\documentclass[twocolumn,aps,prl,superscriptaddress,showpacs,preprintnumbers,amsmath,amssymb]{revtex4}

\usepackage{graphicx}
\usepackage{dcolumn}
\usepackage{bm}

\begin{document}

\title{Resonant speed meter for gravitational wave detection}

\author{Atsushi Nishizawa}
\email{atsushi.nishizawa@nao.ac.jp}
\affiliation{Graduate School of Human and Environmental Studies, Kyoto University, 
Kyoto 606-8501, Japan}
\author{Seiji Kawamura}
\affiliation{TAMA Project, National Astronomical Observatory of Japan, Mitaka, 
Tokyo 181-8588, Japan}
\author{Masa-aki Sakagami}
\affiliation{Graduate School of Human and Environmental Studies, Kyoto University, 
Kyoto 606-8501, Japan}

\date{\today}

\begin{abstract}
Gravitational-wave detectors have been well developed and operated with high sensitivity. However, they still suffer from mirror displacement noise. In this paper, we propose a resonant speed meter, as a displacement noise-canceled configuration based on a ring-shaped synchronous recycling interferometer. The remarkable feature of this interferometer is that, at certain frequencies, gravitational-wave signals are amplified, while displacement noises are not.
\end{abstract}

\pacs{04.80.Nn, 95.55.Ym}
\maketitle

In a gravitational wave (GW) detector using a ground-based laser interferometer, such as LIGO, VIRGO, GEO 600 and TAMA 300 \cite{bib0}, mirror displacement noise, including seismic, thermal, radiation pressure noise, etc., is dominant at low frequencies, and limits the sensitivity. Recently, a design for a displacement-noise-free interferometer (DFI) for GW detection was originally suggested by Kawamura and Chen \cite{bib1} and developed \cite{bib3,bib4}. The idea is based on two principles. One is that GWs and test-mass displacements contribute differently to light propagation time. The mirror motion affects light when the light is reflected by the mirror, while GWs affect light during the propagation of the light. The other is that, for the cancellation of displacement noise, mirror displacement has to be sensed simultaneously and redundantly by independent light beams. According to these two principles, combining each signal between test masses, one can form the displacement-noise-free signal with GW contribution. Therefore, in principle, the sensitivity of the DFI signal is limited only by shot noise. In conventional DFIs, the magnitude of shot noise is as large as that of a Michelson-type interferometer, since the conventional DFIs neither use a cavity nor amplify gravitational-wave signals \cite{bib14}. To further improve the sensitivity, a detector design with a cavity is needed. 

In this paper, we propose such a design, a so-called {\it{resonant speed meter}}. Our reason for calling it a resonant speed meter is described later.
In this configuration, displacement noise is suppressed with the noise cancellation method different from DFI. In DFI, the mirror displacements have to be sensed at the same time with two or more beams, while in this method, they are sensed at different times with the interval $\Delta t$. This means that, at the frequency $(\Delta t)^{-1}$, the phase of the mirror motion rotates by $2\pi$. In other words, the beams experience the mirror displacement in phase. Thus, the mirror displacements at different times can be canceled by subtracting the signals of these beams. Note that this cancellation occurs at the multiple frequencies of $(\Delta t)^{-1}$, and that residual displacement noise remains at other frequencies \cite{bib12}. 

On the other hand, the cavity has to be designed to amplify the GW signals. Such a design is possible if one takes advantage of the quadrupole nature of GWs. At the noise cancellation frequency $(\Delta t)^{-1}$, if one beam propagates in one direction during the first half period and in the right-angled direction during the other half period, the GW signal can survive. For the beam propagating on the same path in the opposite direction, the GW signal with the opposite sign can also survive. Then, subtracting the two beams gives an amplified GW signal. 

\begin{figure}[b]
\begin{center}
\includegraphics[width=6.5cm]{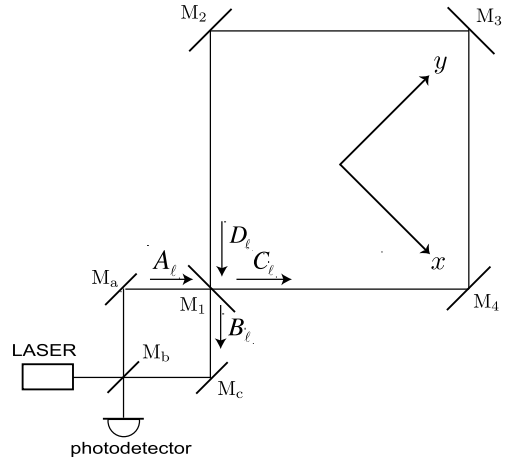}
\caption{Design of a resonant speed meter. A synchronous recycling cavity is composed of the four mirrors $\rm{M}_1-\rm{M}_4$. The electric fields of a CCW beam $A_{\ell}$, $B_{\ell}$, $C_{\ell}$, and $D_{\ell}$ are shown. The fields of CW beam are obtained by reversing along the $y$ axis.}
\label{fig2}
\end{center}
\end{figure}

The detector configuration of the resonant speed meter is shown in Fig. \ref{fig2}. Laser beams divided at the balanced beamsplitter $\rm{M}_b$ are reflected by completely reflective mirrors $\rm{M}_a$ and $\rm{M}_c$, and enter the (ring-shaped) synchronous recycling cavity, which is formed by an input mirror $\rm{M}_1$ and three high-reflective mirrors $\rm{M}_2-\rm{M}_4$. In the cavity, each beam circulates clockwise (CW) and counterclockwise (CCW), then leaves the cavity and is finally recombined at the beamsplitter $\rm{M}_b$ \cite{bib17}. In the presence of GWs, the modulation signals due to GWs at certain frequencies resonate in the cavity. In this paper, we will derive the detector responses to mirror displacements and GWs in the resonant speed meter, and show that, at certain frequencies, the displacement noises can be suppressed, while GW signals are amplified. We also show that the detector sensitivity to GWs can be improved proportional to the circulating number of light in the cavity.



{\it{Detector response}}.---$\rm{M}_1$ has an amplitude reflectivity and transmissivity $(R_F, T_F)$. As for the three other mirrors $\rm{M}_2$, $\rm{M}_3$, $\rm{M}_4$, one can deal with the reflectivities introducing the composite reflectivity $R_E$. For simplicity, none of the mirrors have loss. Let us denote the displacement of $\rm{M}_1$, $\rm{M}_2$, $\rm{M}_3$, and $\rm{M}_4$, in the absence of GWs, as $y_1(t)$, $x_2(t)$, $y_3(t)$, and $x_4(t)$, respectively, where the coordinates $x$ and $y$ are defined in Fig. \ref{fig2}. Electric fields in the cavity are phase-shifted due to the GWs and the displacements of the mirrors. The field circulating CCW (denoted by fixing the subscript $\ell$) can be written as
\begin{eqnarray}
D_{\ell}(t)&=&R_E C_{\ell}(t-4\tau)\; \exp \big[ 4i \omega \tau + i \phi ^{(g)}_{\ell}(t) + \sqrt{2}\,i \omega /c \nonumber \\
&\times & \{ -x_2(t-\tau) +y_3(t-2\tau)+x_4(t-3\tau) \} \big] \;, 
\label{eq1}
\end{eqnarray} 
where $\tau \equiv L/c$, $c$ is the speed of light, $L$ is the side length of the cavity, and $\omega$ is the angular frequency of laser light. $\phi ^{(g)}_{\ell}(t)$ is the phase shift created by the GW and is expressed as the sum of each-side contribution during the propagation in the cavity,
\begin{eqnarray}
\phi ^{(g)}_{\ell}(t)&\equiv &\phi _{21}(t)+\phi _{32}(t-\tau) +\phi _{43}(t-2\tau) 
\nonumber \\ 
&&+\phi _{14}(t-3\tau) \,,
\label{eq9}  
\end{eqnarray}
where, say, $\phi _{21}(t)$ is the phase shift due to the GW during the light trip from $\rm{M}_2$ to $\rm{M}_1$. The junction conditions at $\rm{M}_1$ are 
\begin{eqnarray}
C_{\ell}(t)&=& R_F D_{\ell}(t)\, e^{-\sqrt{2}\, i\, \omega \,y_1(t)/c } +T_F A_{\ell}(t) \;, 
\label{eq2} \\
B_{\ell}(t)&=&T_F D_{\ell}(t) -R_F A_{\ell}(t)\, e^{\sqrt{2}\, i\, \omega\, y_1(t)/c }  \;.
\label{eq3} 
\end{eqnarray}
Equations (\ref{eq1}), (\ref{eq2}), and (\ref{eq3}) are solved separately from the CW fields, and give
\begin{eqnarray}
B_{\ell}(t) &=&  - R_F A_{\ell} (t) \,e^{\sqrt{2}\, i\, \omega\, y_1 (t)} \nonumber \\
&+&\sum_{k=1}^{\infty} T_F^2 R_E^k R_F^{k-1} A_{\ell} (t-4k\tau )\, \nonumber \\
&\times& \exp \bigl[  4 i k \omega \tau +\sqrt{2}\, i\, \omega\, y_1 (t-4k\tau ) \nonumber \\
&+& i \sum _{k^{\prime}=1}^{k} \{\phi ^{(g)}_{\ell}(t-4(k^{\prime}-1)\tau) \nonumber \\
&&\;\;\;\;\;\;\;\;+ \phi ^{(d)}_{\ell}(t-4(k^{\prime}-1)\tau) \} \bigr]\,, \nonumber
\end{eqnarray}
where $\phi ^{(d)}_{\ell}(t)$ is the phase shift due to the displacement of the mirrors and is expressed by
\begin{eqnarray}
\phi ^{(d)}_{\ell}(t) &\equiv & \sqrt{2}\, \omega /c \times \bigl[ -x_2(t-\tau ) + y_3(t-2\tau )\nonumber \\
&&+x_4(t-3\tau ) -y_1(t-4\tau ) \bigr]\,. 
\label{eq10}
\end{eqnarray}
We assume that the carrier field at $\rm{M}_1$ is $A_{\ell}(t)=A_0 e^{i\omega t}$ and nonperturbed cavity length satisfies the resonant condition $4\omega \tau = 2\pi n,\,n=1,2,\cdots$. In addition, $\phi_{\ell}^{(g)} \ll 1$, $\phi_{\ell}^{(d)} \ll 1$, and $\omega\, y_1 /c \ll 1$ are assumed. Using these assumptions and defining 
${\cal{T}}_{\ell}(t) \equiv B_{\ell}(t)/A_{\ell}(t)$, we obtain
\begin{eqnarray}
{\cal{T}}_{\ell}(t) &\approx & -R_F \biggl[ 1+\sqrt{2} i\, \omega\, y_1(t) \nonumber \\
&&- T_F^2 \sum_{k=1}^{\infty} R_E^k R_F^{k-2} \bigl[ 1+\sqrt{2} i\, \omega\, y_1 (t-4k\tau ) \nonumber \\
&&+ i \sum _{k^{\prime}=1}^{k} \{ \phi ^{(g)}_{\ell} + \phi ^{(d)}_{\ell} \} (t-4(k^{\prime}-1)) \bigr]  \biggr]\,. 
\label{eq4}
\end{eqnarray}
Fourier transformation of Eq. (\ref{eq4}), it can be written in the form,
\begin{eqnarray}
\tilde{{\cal{T}}}_{\ell}(\Omega) &=& \sqrt{2}\, i\, \omega \beta (\Omega) \tilde{y_1} + i \alpha^{\prime} (\Omega ) (\tilde{\phi} ^{(g)}_{\ell} +\tilde{\phi} ^{(d)}_{\ell}) \,, 
\label{eq5} \\
\beta (\Omega) &\equiv & \frac{-R_F+R_E e^{-4i \Omega \tau}}
{1-R_F R_E e^{-4i \Omega \tau}}\,, \nonumber \\
\alpha^{\prime}(\Omega ) &\equiv & \frac{T_F^2 R_E}
{(1-R_F R_E)(1-R_F R_E e^{-4i \Omega \tau})}\,. \nonumber
\end{eqnarray}
Here $\beta(\Omega)$ and $\alpha^{\prime}(\Omega)$ are optical amplification factors due to the cavity. According to the axisymmetry of the system along the $y$ axis, the transfer function for the CW beam is obtained by replacing the subscripts, $4\rightarrow 2$ and $2\rightarrow 4$, and reversing the sign of $\tilde{x}$, then,
\begin{equation}
\tilde{{\cal{T}}}_{r}(\Omega) = \sqrt{2}\, i\, \omega \beta (\Omega) \tilde{y_1} + i \alpha^{\prime} (\Omega ) (\tilde{ \phi} ^{(g)}_{r} +\tilde{ \phi} ^{(d)}_{r}).
\label{eq6} 
\end{equation}

The differential of the two beams is detected at the photodetector, and gives
\begin{eqnarray}
\tilde{{\cal{T}}}(\Omega) &\equiv &\tilde{{\cal{T}}}_{r}(\Omega ) -\tilde{{\cal{T}}}_{\ell} (\Omega) \nonumber \\
&=& -i \alpha ^{\prime} \bigl[ (\tilde{ \phi} ^{(g)}_{r} -\tilde{ \phi} ^{(g)}_{\ell}) +(\tilde{ \phi} ^{(d)}_{r} -\tilde{ \phi} ^{(d)}_{\ell}) \bigr]\,, \nonumber \\
\tilde{ \phi} ^{(d)}_{r} -\tilde{ \phi} ^{(d)}_{\ell} &=& 2 \sqrt{2}\, i\, \omega /c \times e^{-2i\Omega \tau} \sin \Omega \tau \, (\tilde{x}_2+\tilde{x}_4)\,. \label{eq7} 
\end{eqnarray}
At this stage, the displacements $\tilde{y}_1$ and $\tilde{y}_3$ in Eqs. (\ref{eq5}) and (\ref{eq6}) are automatically canceled out at any frequencies. This is because the CW and CCW beams simultaneously experience the displacement of $\rm{M}_1$ and $\rm{M}_3$ \footnote{The arrival times of light at $\rm{M}_1$, or $\rm{M}_3$ depend on the displacement of $\rm{M}_2$ and $\rm{M}_4$. However, it is the order of ${\cal{O}}(\tilde{x}^2/L^2)$ and can be ignored.}. In addition, at the frequencies that satisfy $\Omega \tau = n \pi, \; n=1,2,\cdots$, Eq. (\ref{eq7}) gives exactly zero, and all displacement noises vanish. This is because the CW and CCW beam in the cavity experience the displacement of $\rm{M}_2$ and $\rm{M}_4$ in phase, though the time of reflection is shifted by multiples of the period. Therefore, in our detector, displacement noises in the cavity are not amplified around the cancellation frequencies, though the cavity is on resonance. 

Suppose that a GW propagates in the direction vertical to the detector plane with the polarization along the direction $\rm{M}_1$-$\rm{M}_2$ and $\rm{M}_1$-$\rm{M}_4$. In the transverse-traceless gauge, it can be written as
\begin{equation}
\mathbf{h}^{TT}(t, z)= h (t-z/c) [ \mathbf{u} \otimes \mathbf{u} -\mathbf{v} \otimes \mathbf{v}] \,, \nonumber
\end{equation}
where $\mathbf{u}$ and $\mathbf{v}$ are unit vectors directed from $\rm{M}_1$ toward $\rm{M}_2$ and $\rm{M}_4$, respectively. Let the detector be on the $x-y$ plane and set $z=0$, for simplicity. The GW-induced phase shift of light during the trip from $\rm{M}_i$ to $\rm{M}_j$ is given by \cite{bib3}
\begin{equation}
\tilde{\phi}_{ij} (\Omega)= \pm \tilde{h} \omega / \Omega \, \times e^{-i\Omega \tau /2}\, \sin(\Omega \tau /2). 
\label{eq11}
\end{equation}
The positive and negative signs correspond to the vertical and horizontal propagations in Fig. \ref{fig2}. $\tilde{h}$ is the Fourier component of GW amplitude. From Eq. (\ref{eq9}) and the counterpart for the CW beam, the Fourier component of the total GW response is 
\begin{eqnarray}
\tilde{ \phi} ^{(g)}_{r} -\tilde{ \phi} ^{(g)}_{\ell} &=& 8 i \,\tilde{h} \omega / \Omega \, \times e^{-2i\Omega \tau} \cos (\Omega \tau) \,\sin ^2(\Omega \tau/2)\;. \nonumber \\
&& \label{eq15}
\end{eqnarray}

\begin{figure}[t]
\begin{center}
\includegraphics[width=7.3cm]{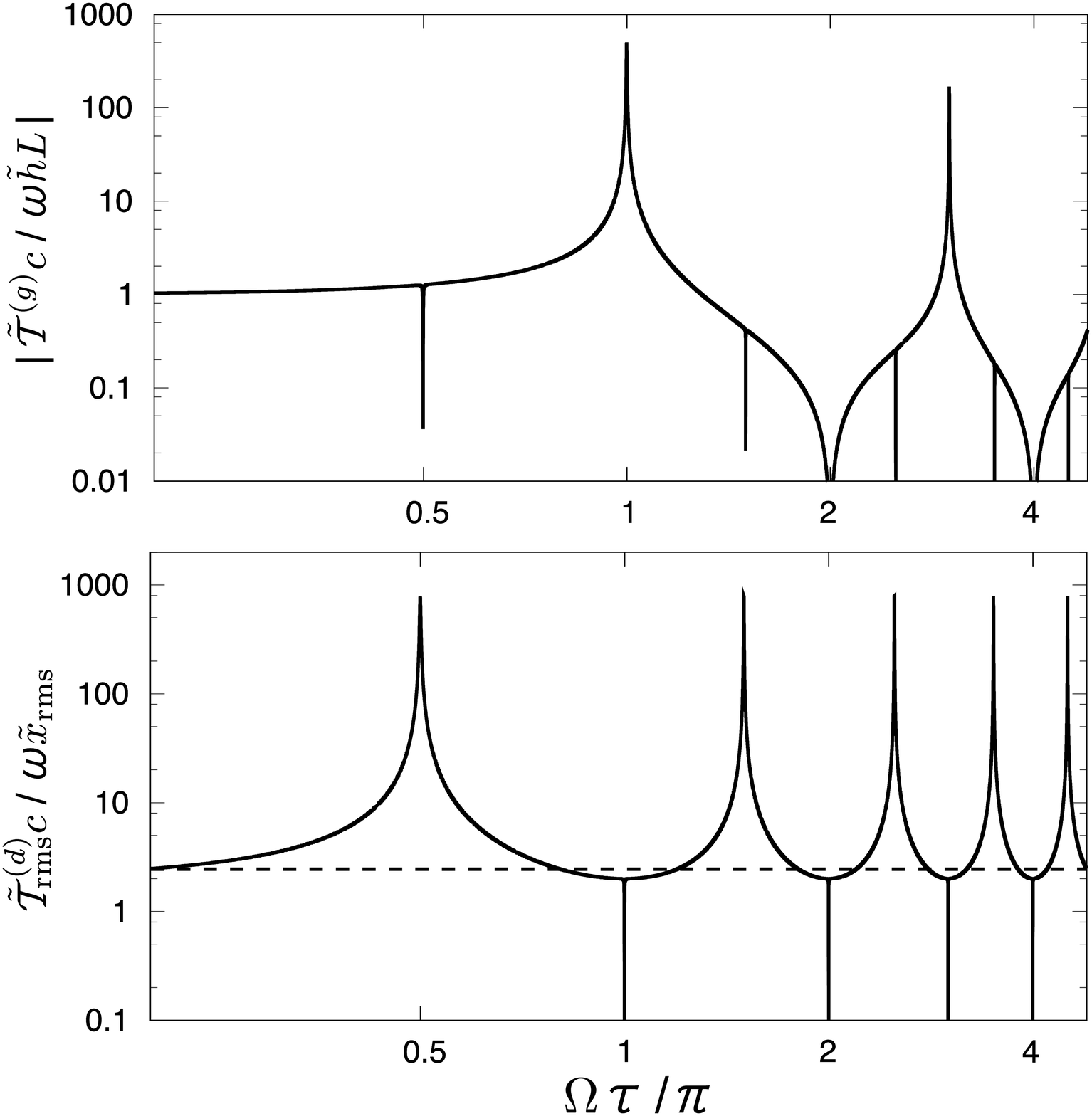}
\caption{GW response $\tilde{\cal{T}}^{(g)}_{\rm{rms}}(\Omega)$ (upper figure), and displacement responses $\tilde{\cal{T}}^{(d)}_{\rm{rms}}(\Omega)$ and $\tilde{\cal{T}}^{(s)}_{\rm{rms}}(\Omega)$ (solid and dashed lines in the lower figure, respectively) as a function of normalized frequency $\Omega \tau / \pi$. Each quantity is normalized to be dimensionless. The reflectivities of mirrors are $R_F=0.99$ and $R_E=1$ for the illustration, which corresponds to $\alpha^{\prime} \approx 200$.}
\label{fig3}
\end{center}
\end{figure}

So far, we have ignored the contribution of displacements at the mirrors in the Sagnac part, ${\rm{M}}_a$, ${\rm{M}}_b$, and ${\rm{M}}_c$, in Fig. \ref{fig2}. In fact, these displacements are not canceled, and will contribute as residual noise. The contribution is
\begin{equation}
\tilde{ \phi}_{r}^{(s)} -\tilde{ \phi}_{\ell}^{(s)} = - \sqrt{2} \omega /c \;(\tilde{x}_a -\tilde{x}_b +\tilde{x}_c),
\end{equation}
where $\tilde{x}_a$, $\tilde{x}_b$, and $\tilde{x}_c$ are Fourier components of the displacements of the mirrors, $\rm{M}_a$, $\rm{M}_b$, and $\rm{M}_c$.

Let us define each contribution of the total detector output as
\begin{eqnarray}
\tilde{\cal{T}}^{(g)} (\Omega)&\equiv & i \alpha ^{\prime} (\tilde{\phi} ^{(g)}_{r} -\tilde{\phi} ^{(g)}_{\ell}), \\
\tilde{\cal{T}}^{(d)} (\Omega)&\equiv & i \alpha ^{\prime} (\tilde{\phi} ^{(d)}_{r} -\tilde{\phi} ^{(d)}_{\ell}), 
\label{eq12} \\
\tilde{\cal{T}}^{(s)} (\Omega)&\equiv & \tilde{ \phi}_{r}^{(s)} -\tilde{ \phi}_{\ell}^{(s)}.
\label{eq13}
\end{eqnarray}
Assuming that the magnitude of disturbances at each mirror is the same (the phase is not), we define $\tilde{x}_{\rm{rms}} = \sqrt{ \langle |\tilde{x}_i |^2 \rangle}$, where $i=1, 2, 3, 4, a, b, c$, and $\langle \cdots \rangle$ denotes the ensemble average. Then, root mean squares of Eqs. (\ref{eq12}) and (\ref{eq13}) are
\begin{eqnarray}
\tilde{\cal{T}}^{(d)}_{\rm{rms}} (\Omega)&=& 4 \omega /c \times |\alpha ^{\prime}\,\sin \Omega \tau | \, \tilde{x}_{\rm{rms}}\,,
\label{eq14} \\
\tilde{\cal{T}}_{\rm{rms}}^{(s)} (\Omega)&=& \sqrt{6}\, \omega /c \times \tilde{x}_{\rm{rms}}\,. 
\end{eqnarray}
The GW response $\tilde{\cal{T}}^{(g)}$, and the displacement responses $\tilde{\cal{T}}^{(d)}_{\rm{rms}}$ and $\tilde{\cal{T}}^{(s)}_{\rm{rms}}$, are shown in Fig. \ref{fig3} as a function of normalized frequency $\Omega \tau / \pi$. The GW signal resonates at $\Omega_{\rm{GW}} = 2\pi \times (2m-1)/ 2\tau$, while the cancellation of the displacement noise in the cavity occurs at $\Omega_{\rm{cancel}} =2\pi \times n / 2\tau$, where $m,n=1,2,\cdots$ \cite{bib15}. For $m=n=1$, the GW signal is amplified, while the cavity displacement noise is not. This can be explicitly seen from Eqs. (\ref{eq7}) and (\ref{eq15}). The degeneracy between the noise cancellation frequencies and the GW-signal amplification frequencies is broken. Thus, there exist the frequencies where the GW signal is amplified and the displacement noise is canceled. 

The noise cancellation mechanism is the same as that of a speed meter \cite{bib13}. This can be understood explicitly in a time domain. From Eq. (\ref{eq10}) and the CW counterpart, the round-trip displacement response in the cavity around the cancellation frequencies can be written as
\begin{eqnarray}
\phi_{r}^{(d)}(t)-\phi_{\ell}^{(d)}(t)&=& \sqrt{2}\, \omega /c \; [ x_2(t-\tau) - x_2(t-3\tau) \nonumber \\
&&+x_4(t-\tau) -x_4(t-3\tau) ] \nonumber \\
&\sim & 2\sqrt{2}\, \omega \tau /c\;  \,[v_2(t-2\tau) +v_4(t-2\tau)]\,. \nonumber
\end{eqnarray}
Our detector senses not the mirror positions but the mirror velocities. Moreover, the GW signal is amplified by the resonant cavity. This is the reason why we call it a resonant speed meter.

\begin{figure}[t]
\begin{center}
\includegraphics[width=7cm]{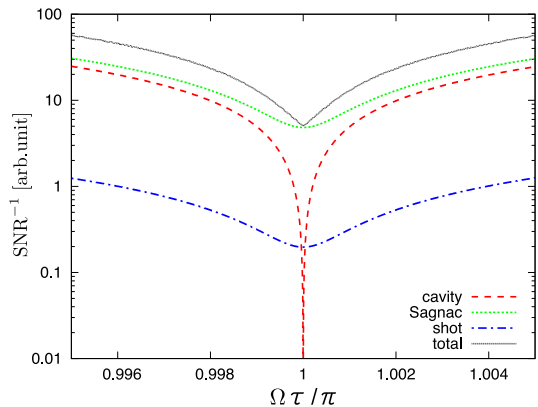}
\caption{(Color online). Relative noise curves of a resonant speed meter. Each curve,  "cavity", "Sagnac", and "shot" is the plot of the displacement noise in the cavity, the displacement noise in the Sagnac part, and quantum (shot) noise, respectively. The curve "total" is the sum of these three noises. The reflectivities are $R_F=0.99$ and $R_E=1$ ($\alpha^{\prime} \approx 200$ at $\Omega \tau / \pi =1$).}
\label{fig5}
\end{center}
\end{figure}

{\it{Sensitivity}}.---The signal-to-noise ratio (SNR) can be calculated by
\begin{equation}
{\rm{SNR}} = \tilde{\cal{T}}^{(g)}\, /\,(\tilde{\cal{T}}^{(d)}_{\rm{rms}} + \tilde{\cal{T}}_{\rm{rms}}^{(s)} + \tilde{\cal{T}}^{(q)}). \nonumber
\end{equation}
Here we added quantum noise $\tilde{\cal{T}}^{(q)}$, which has a frequency-independent spectrum. In Fig. \ref{fig5}, the noise curve of each component is shown. For the illustration of the noise cancellation, we choose the displacement noise significantly dominant, namely, $\tilde{x}_{\rm{rms}} = 10\, \tilde{x}_{\rm{shot}}$ at $\Omega \tau /\pi =1$, where $\tilde{x}_{\rm{shot}}\equiv \tilde{\cal{T}}^{(q)}\,c/\omega$ (Note that whether the displacement noise is dominant or not depends on the detection frequencies.). Around $\Omega \tau /\pi =1$, cavity displacement noise is canceled. The cancellation does not exactly occur at other frequencies and the residual noise creates a dip on the displacement noise spectrum at the cancellation frequencies. The bandwidth of the dip is determined by the optical path length between $\rm{M}_2$ and $\rm{M}_4$, in other words, by the optical time lag that CW or CCW beams experience the displacements of $\rm{M}_2$ and $\rm{M}_4$. The time lag is $2\tau$ in our detector and gives the factor $\sin \Omega \tau$ in Eq. (\ref{eq14}). The total sensitivity at $\Omega \tau /\pi =1$ is limited by residual displacement noise in the Sagnac part (or shot noise if the displacement noise is relatively small). However, both noises decrease proportional to $\alpha^{\prime}$ because of the amplification of the GW signal. On the other hand, the displacement noise of the cavity mirrors is independent of $\alpha^{\prime}$, but is already canceled around $\Omega \tau / \pi =1$. Thus, within the narrow bandwidth, the total noise level diminishes as $\alpha^{\prime}$ increases. 

\begin{figure}[t]
\begin{center}
\includegraphics[width=7.5cm]{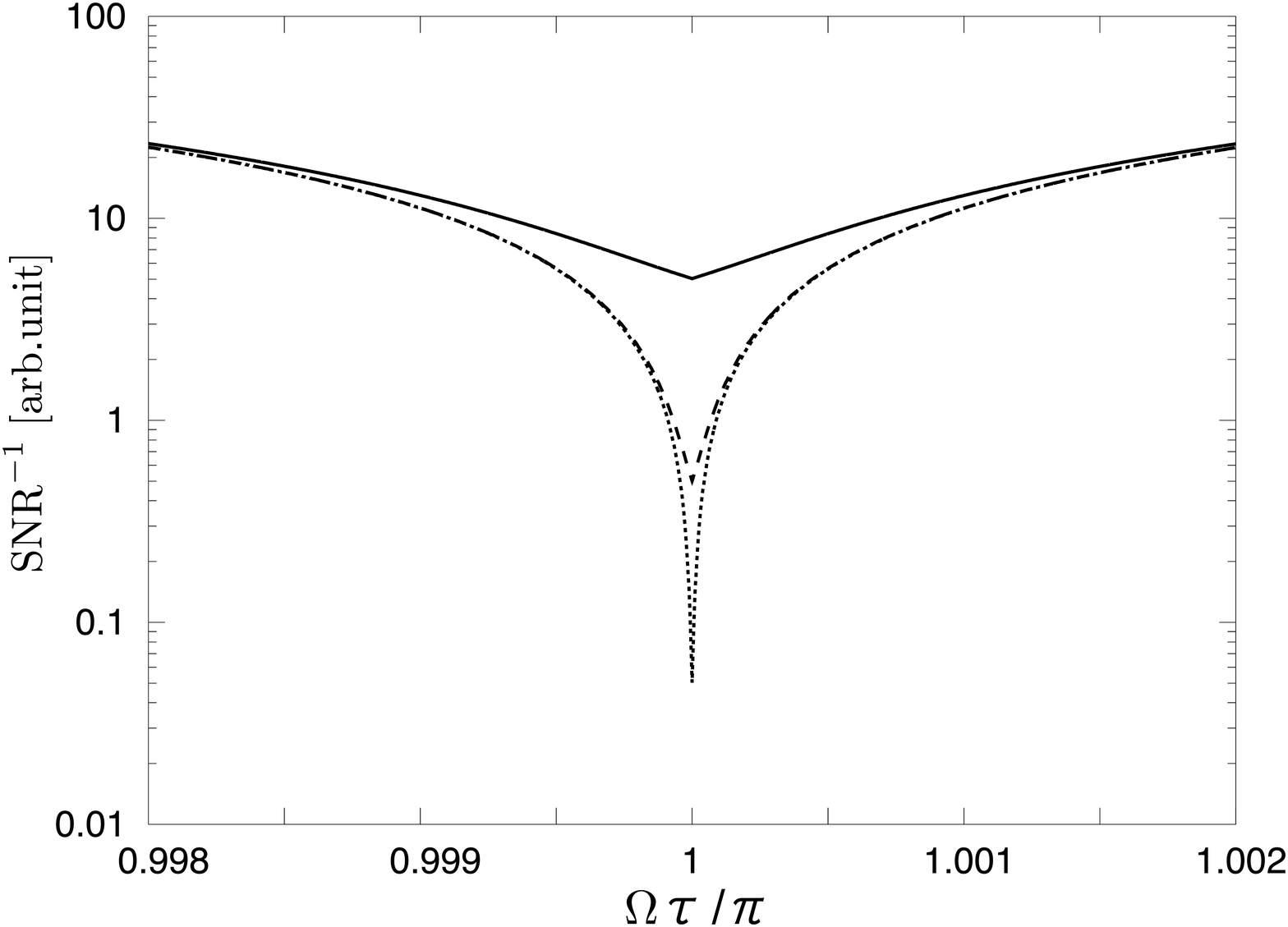}
\caption{Dependence of relative total-noise curve on $\alpha^{\prime}$. Solid, dashed, and dotted curves are when $\alpha^{\prime} \approx 200, 2000, 20000$ at $\Omega \tau / \pi =1$, corresponding to $R_F=0.99, 0.999, 0.9999$, respectively, and $R_E=1$.}
\label{fig6}
\end{center}
\end{figure}

In Fig. \ref{fig6}, the dependence of the relative total noise curve on $\alpha^{\prime}$ is shown. The noise cancellation allows us to improve the sensitivity, being proportional to $\alpha^{\prime}$ without being limited by displacement noises. Roughly speaking, the bandwidth of the total noise curve is inversely proportional to $\alpha^{\prime}$.  

{\it{Conclusions}}.---In this paper, we suggested a resonant speed meter for GW detection. In general, the displacement noise in the cavity is amplified at certain frequencies along with GW signals. However, in our detector configuration with a ring-shaped synchronous recycling cavity, the displacement noises are suppressed, while GW signals are amplified. The sensitivity is improved proportional to the circulating number of light in the cavity within a narrow bandwidth. Such a detector is considerably effective for detecting GW backgrounds and could be one of the alternatives to resonant-bar-like detectors in the future.

We thank K. Somiya, Y. Chen, and T. Akutsu for helpful comments and discussions. We also acknowledge B. Lantz for the careful reading the paper and giving us useful comments. This research was partially supported by the Ministry of Education, Science, Sports and Culture, Grant-in-Aid for Scientific Research (A), 17204018, and the Mitsubishi Foundation. We also would like to acknowledge the U.S. National Science Foundation.

\bibliography{SR-DFI}

\begin{thebibliography}{11}
\bibitem[{bib({\natexlab{c}})}]{bib12}
\bibinfo{note}{This property is the same as that of the Sagnac interferometer,
  though a cavity is not used therein. See, for example, ref. \cite{bib11}.}
\bibitem[{bib({\natexlab{d}})}]{bib17}
\bibinfo{note}{This design is well known as a laser-ring gyroscope. It picks up
  earth rotation and shifts the frequency of the counterpropagating beams,
  which is called Sagnac effect. The effect can be ignored when the size of the
  detector is relatively small. However, for a large detector, some measures
  must be done to compensate the phase shift.}
\bibitem[{bib({\natexlab{e}})}]{bib15}
\bibinfo{note}{In a conventional synchronous recycling interferometer, which is
  an L-shaped cavity version of a resonant speed meter, $\Omega_{\rm{GW}} =
  2\pi \times (2m-1)/ 4\tau$ and $\Omega_{\rm{cancel}} = 2\pi \times 2n /
  4\tau$. No solution exists. See ref. \cite{bib16} in detail.}
\end{thebibliography}

\end{document}